# Pressure Induced Fermi Surface Deformation in Lithium

A.Rodriguez-Prieto[&] and A. Bergara[*]

*Materia Kondentsatuaren Fisika Saila, Zientzia eta Teknologia Fakultatea,*

*Euskal Herriko Unibertsitatea, 644 Postakutxatila, 48080 Bilbo, Basque Country, Spain*

and

*Donostia International Physics Center (DIPC),*

*Paseo de Manuel Lardizabal, 20018, Donostia, Basque Country, Spain*

[&]alvaro@wm.lc.ehu.es

[*]a.bergara@lg.ehu.es

## I. Summary

Recently reported structural complexity and superconducting transition in lithium under pressure has increased the interest in light alkalis, otherwise considered as simple and well known systems under normal conditions. In this work we present an analysis of the pressure induced Fermi surface deformation in lithium with increasing pressure. According to our calculations, under pressure the Fermi surface becomes highly anisotropic and around 30 GPa develops an extended nesting, which could be the origin of the complex phase transitions observed at this pressure via a phonon instability. On the other hand, the phonon softening induced by the observed distortion and nesting in the Fermi surface could also imply an increasing electron-phonon interaction with pressure, which might also allow a better understanding of the observed superconducting transition in lithium at around the same pressure range.

## I. Introduction

Light alkali metals have usually been considered *simple* metals due to their monovalency, high conductivity and crystallization in high symmetric structures. Under normal conditions the interaction between valence electrons and ionic cores is weak in these elements and the simple nearly free electron (NFE) model has been considered a good approximation to analyze their properties (Wigner and Seitz 1933), which can be easily checked by their almost spherical Fermi surfaces. However, recent both theoretical and experimental results have radically changed our picture of the alkalis as simple systems, as phase transitions to low coordinated structures induced by electronic localization have been reported (Neaton and Ashcroft 1999, Hanfland *et al.* 2000, Christensen and Novikov 2001, Bergara *et al.* 2001). More surprisingly, recent experiments have shown that lithium even superconducts at around 20 K when the applied pressure rises to 30 GPa (Shimizu *et al.* 2002, Struzhkin *et al.* 2002, Deemyad and Schilling 2004), becoming the highest transition temperature between simple elements (Ashcroft 2002). It is noteworthy that experiments looking for superconductivity in lithium under ambient pressure have failed (Juntunen and Tuoriniemi 2004), which even rises the interest to characterize physical properties of compressed light elements.

In what follows, we present an *ab initio* analysis of the pressure induced deformation of the Fermi surface in lithium. According to our calculations, the Fermi surface develops a nesting under pressure, which could be the origin of the interesting and unexpected new properties arising in compressed lithium. Section III describes the theoretical and computational method that we have used in this study. In Sec. IV we present our results and, finally, conclusions are presented in Sec. V.

### III. Theoretical and Computational Method.

In order to analyze the electronic properties of lithium under pressure we have used a plane-wave implementation of Density Functional Theory (DFT) (Hohenberg and Kohn 1964, Kohn and Sham 1965) within the Local Density Approximation (LDA) (Perdew and Wang 1992), as implemented in the Vienna Ab Initio Simulation Package (VASP) (Kresse and Hafner 1993, Kresse and Furthmuller 1996). When dealing with high pressures an important core overlap is observed and requires methods that actually take into account the core electrons, instead of just adding them to the ionic cores. Therefore, we have considered the Projector Augmented Wave (PAW) approximation (Blochl 1994), which fully treats all the electrons. In order to obtain an accurate description of the whole Fermi surface, a dense Monkhorst-Pack mesh of 40 x 40 x 40 has been considered and three dimensional graphics have been generated by XCrySDen (Kokalj 1999). On the other hand, for each cross section of the Fermi surface we have used a different set of around 3000 k-points in the corresponding plane inside the Brillouin zone, with a denser mesh in the proximities of the Fermi line. The description of the ML is implemented by a supercell which, in order to minimice interactions between them, contains ten layers of vacuum between MLs. The energy cutoff for all the calculations carried out is 1090 eV.

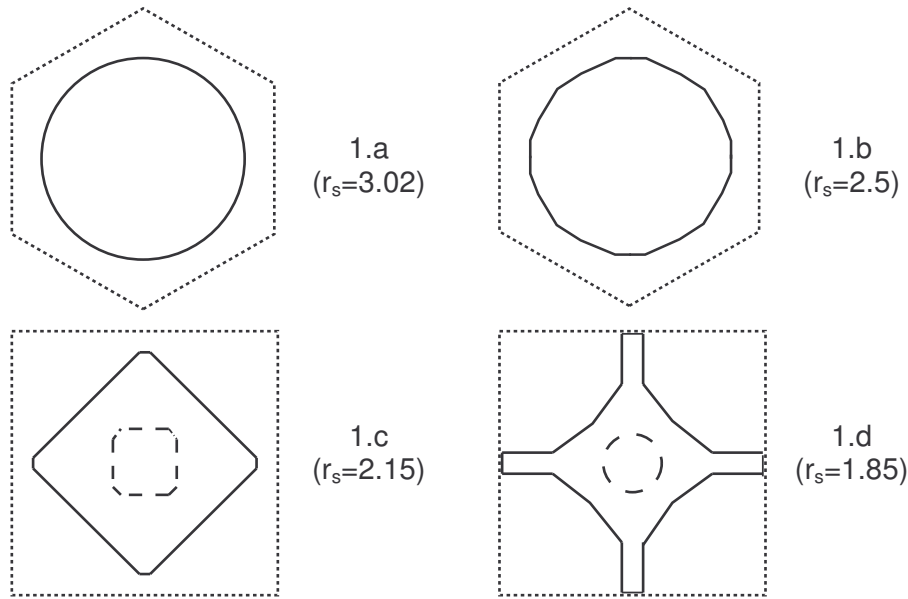

**Fig.1: Fermi *lines* of a monoatomic hexagonal lithium ML at equilibrium, $r_s$=3.02, (a) and $r_s$=2.5 (b), and a monoatomic square lithium ML at $r_s$=2.15 (c) and $r_s$=1.85 (d). Solid lines correspond to the Fermi *line* of the *s*-like band and, as at $r_s$=2.4 the $p_z$-like band starts to be occupied, the Fermi *line* associated to this band is represented by a dashed line. Dotted lines represent the Brillouin zone boundary. The circular shape observed at $r_s$=3.02, which shows a NFE-like behavior at equilibrium, significantly breaks under compression. Particularly, at $r_s$=2.15 the Fermi *line* becomes a perfect square, as corresponds to a half filled tight binding-like model, indicating a strong electronic coupling in the *ΓM* direction. Although at higher densities the observed nesting in the Fermi line breaks, it is noteworthy that around $r_s$=2.15 it induces phase transitions to more complex structures (Rodriguez-Prieto and Bergara 2005a).**

## IV. Results.

An essential feature determining electronic properties of materials is the geometry of the Fermi surface. In this section we present the analysis of the pressure induced modification of the Fermi surface in both a lithium monolayer (ML) and bulk. Besides the intrinsic interest to analyze physical properties of systems with reduced dimensionality, our main motivation to study a ML is related to its simpler geometrical configuration, which allows us to make a deeper analysis of the interesting new physical properties that can be induced by pressure, and will be used as well as a first guide for a better understanding of the compressed bulk lithium.

### a. Li Monolayer:

Lithium ML at equilibrium, $r_s$=3.02 ($r_s$ is the two-dimensional linear density parameter, defined by the relation $A/N=\pi(r_s a_0)^2$ where $N$ is the number of nuclei in a ML with area $A$ and $a_0$ is the Bohr radius), favors the compact monoatomic hexagonal structure. However, at $r_s < 2.25$ the more open monoatomic square structure becomes favored over the hexagonal one (Bergara *et al.* 2001). As it is shown in Fig. 1a, at equilibrium just one Fermi *line* exists, associated to the *s* band, which is almost circular as corresponds to a NFE approximation. However, this simple picture breaks under pressure. For example, at higher densities ($r_s$<2.4) the second band, of $p_z$ character, starts to be occupied around the $\Gamma$ point. Therefore, at $r_s$=2.15 there are two Fermi *lines* associated to the two occupied *s* and $p_z$ bands and, surprisingly, the Fermi *line* associated to the *s* band becomes a perfect square as corresponds to a half filled tight binding-like model (Fig. 1c). As it is well known, this model becomes appropriate to describe electronic properties of systems with localized electronic states, just being the opposite case of the NFE approximation valid at ambient pressures. The perfect nesting observed at $r_s$=2.15 strongly couples electronic states close to the Fermi *line* along the $\Gamma M$ direction, and becomes the physical origin of the proposed phase transitions to more complex structures at higher densities in the ML (Rodriguez-Prieto and Bergara, 2005a).

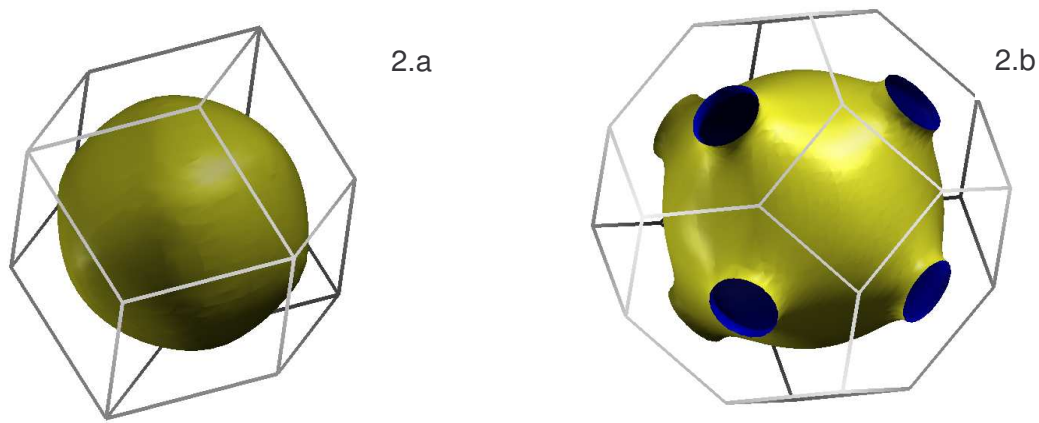

**Fig.2: Fermi surfaces of *bcc*-lithium at equilibrium (a) and *fcc*-lithium at P=30 GPa (b). The almost perfect sphere obtained at equilibrium confirms the validity of the NFE approximation for lithium at ambient pressures. However, under compression the sphere becomes strongly distorted, with necks along the $\Gamma L$ direction and even presents parallel faces along the $\Gamma K$ direction.**

*b. Li Bulk:*

At normal conditions of pressure and temperature bulk lithium favors the *bcc* structure, but with a slight energy difference over the *fcc*. However, when the applied pressure rises to around 8 GPa the *fcc* structure emerges as the preferred one and it remains favored up to around 40 GPa, where new more complex structures (i.e. *cI16*) starts to be favored (Hanfland *et al.* 2000).

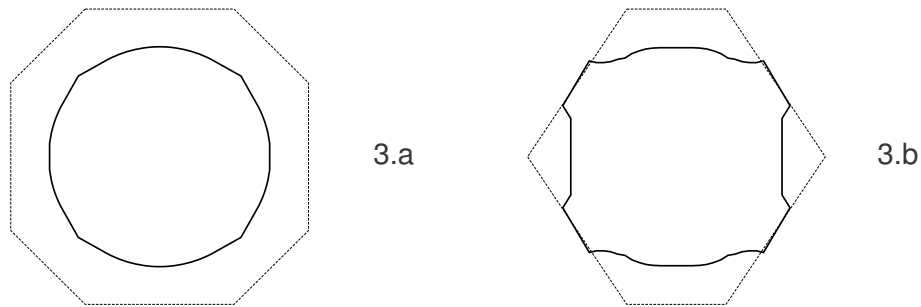

**Fig.3: Fermi surface cross sections (solid lines) of *fcc*-lithium at 30 GPa along $\Gamma KX$ (a) and $\Gamma KL$ (b) planes. Dotted lines represent the Brillouin zone boundary. Although the cross section along the $\Gamma KX$ plane shows an almost perfect circle, it becomes highly distorted in the $\Gamma KL$ plane with a clear nesting along the $\Gamma K$ direction.**

Fig. 2 displays the Fermi surface of *bcc*-lithium at equilibrium (Fig. 2a) and *fcc*-lithium at P=30 GPa (Fig. 2b). The almost perfect sphere obtained at ambient pressure deforms significantly under compression, losing its spherical shape and, similar to the Fermi surface of cooper at normal conditions, presents necks along the $\Gamma L$ direction. More interestingly, as will be described bellow, we can also distinguish almost parallel faces connecting each two nearest neighbor necks. Fig. 3a and 3b show cross sections of the Fermi surface along (100) and (110) planes ($\Gamma KX$ and $\Gamma KL$ planes), respectively. As a result of the increasing non-local character of the pseudopotential (Bergara *et al.* 2001) the Fermi surface of lithium under pressure becomes highly anisotropic and, although the cross section along the $\Gamma KX$ plane shows an almost perfect circle, it becomes highly distorted in the $\Gamma KL$ plane and a clear nesting in the $\Gamma K$ direction (equivalent to the $\Gamma M$ direction in the ML) can be also observed.

In order to analyze the extension and magnitude of the nesting, Figs. 4a-4d show cross sections of the Fermi surface of *fcc*-lithium at 30 GPa along planes parallel to $\Gamma KL$ at 0.2, 0.3, 0.4 and 0.6 times the $\Gamma K$ distance, respectively. It is interesting to point out that the nesting persists in the parallel planes and, although the magnitude of the nesting momentum slightly decreases as the cross section plane approaches the zone boundary, its weight increases because the Fermi surface area connected by the nesting becomes larger. This Fermi surface anisotropy is expected to be the origin of important physical properties in lithium at this pressure range. For example, our preliminary calculations indicate that at around 30 GPa a phonon mode becomes unstable, associated to the Kohn anomaly corresponding to the observed nesting (Rodriguez-Prieto and Bergara, 2005b), which could explain the complex phase transitions observed in this pressure range.

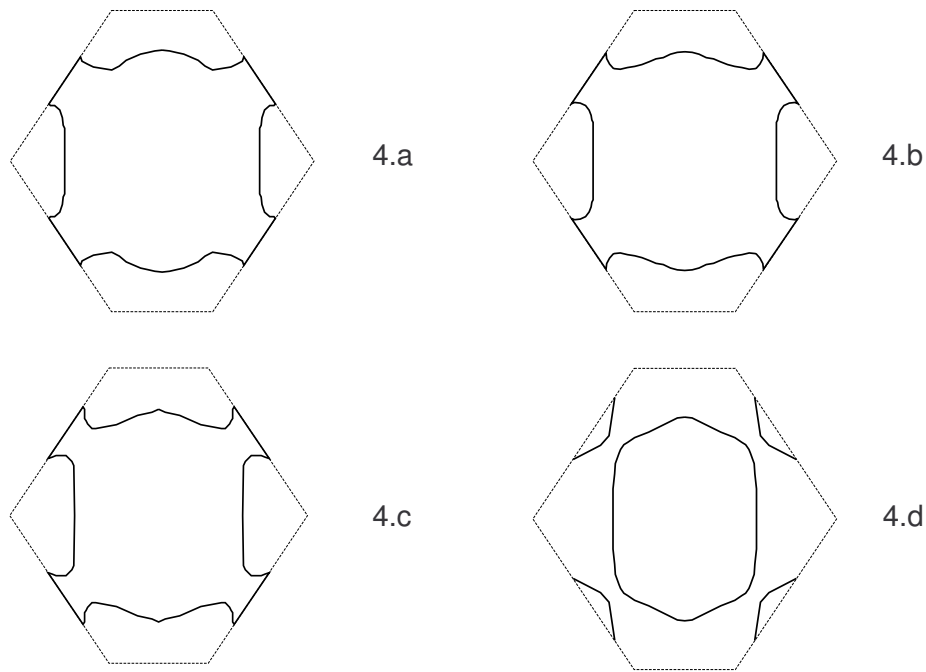

**Fig.4: Fermi surface cross sections (solid lines) of *fcc*-lithium at 30 GPa along planes parallel to the *ΓKL* plane at 0.2 (a), 0.3 (b), 0.4 (c) and 0.6 (d) times the *ΓK* distance. Dotted lines represent the Brillouin zone boundary.**

## V. Conclusions.

We have performed *ab initio* calculations of the Fermi surface deformation with increasing pressure in both lithium ML and bulk. According to our calculations, although the nearly free electron model correctly describes the properties of lithium at normal conditions, it loses its validity as the electronic density increases. As a result of the increasing non-local character of the pseudopotential, the Fermi surface becomes highly anisotropic under pressure and at 30 GPa even shows an extended nesting along the *ΓK* direction. Preliminary calculations indicate that the observed Fermi surface deformation induces a phonon instability associated to the nesting momentum vector (Rodriguez-Prieto and Bergara, 2005b), which can be the origin of the observed complex phase transitions in the same pressure range. On the other hand, related to the observed phonon softening the electron-phonon interaction might also become large in lithium under pressure, so that an analysis based on the evolution of the Fermi surface might also provide a new perspective to understand the experimentally observed high superconducting transition temperature.

## VI. Bibliography.